\documentclass[aps,prb,twocolumn,superscriptaddress,floatfix,longbibliography]{revtex4-2}
\usepackage[utf8]{inputenc}
\usepackage[T1]{fontenc}
\usepackage{amsmath, amssymb, bm}
\usepackage{graphicx}
\usepackage{xcolor}
\usepackage{mathtools}
\usepackage{hyperref}
\usepackage{cleveref}
\usepackage{booktabs}
\usepackage{algorithmic}
\usepackage{physics}
\usepackage{siunitx}

\hypersetup{
    colorlinks,
    linkcolor={blue!50!black},
    citecolor={blue!50!black},
    urlcolor={blue!80!black},
}

\renewcommand{\vec}[1]{{\mathbf{\boldsymbol{#1}}}}
\renewcommand{\vr}{\mathbf{r}}
\newcommand{\vk}{\mathbf{k}}
\newcommand{\vK}{\mathbf{K}}
\newcommand{\vh}{\mathbf{h}}

\newcommand{\MLP}{\operatorname{MLP}}

\begin{document}

\title{Fast, Accurate, and Scalable Fermionic Neural Networks via Translation Equivariance}

\author{David D. Dai}
\email{dddai@mit.edu}
\thanks{Part of this work was completed while interning at Taiwan Semiconductor Manufacturing Company (TSMC).}
\affiliation{Department of Physics, Massachusetts Institute of Technology, Cambridge, Massachusetts 02139, USA}
\affiliation{The NSF AI Institute for Artificial Intelligence and Fundamental Interactions}
\affiliation{R\&D, Taiwan Semiconductor Manufacturing Company, Hsinchu, Taiwan}
\author{Yen-Ting Lin}
\affiliation{R\&D, Taiwan Semiconductor Manufacturing Company, Hsinchu, Taiwan}
\author{Marin Soljačić}
\email{soljacic@mit.edu}
\affiliation{Department of Physics, Massachusetts Institute of Technology, Cambridge, Massachusetts 02139, USA}
\affiliation{The NSF AI Institute for Artificial Intelligence and Fundamental Interactions}

\date{\today}

\begin{abstract}
We demonstrate that designing a neural quantum state to be an exact eigenstate of the Hamiltonian's symmetries significantly improves both training speed and final variational energy.
For the 2D electron gas, we design TorFormer, a neural network wavefunction which is an exact eigenstate of the total momentum.
TorFormer describes both the Fermi liquid and Wigner crystal with no supervision and significantly outperforms Psiformer-based references up to large system sizes.
For $r_s = 30.0$ and $40.0$ at $N=91$, we compare TorFormer trained for $8\mathrm{K}$ steps against the previous best NQS, which required $100\mathrm{K}$ training steps.
Our improvement to the total energy at $r_s = 40.0$, excluding the trivial Madelung part, is $0.12\%$---enormous compared to the tiny differences separating phases.
Relative to Slater-Jastrow-backflow diffusion Monte Carlo, TorFormer's energy decrease is roughly $\num{9.8}$ times that of the previous best NQS.
Our work demonstrates that neural quantum states can both accurately and efficiently solve large-scale problems.
\end{abstract}

\maketitle

\section{Introduction}

    Neural quantum states---neural networks designed to represent many-body wavefunctions---have emerged as a powerful new tool for tackling the many-electron Schr\"odinger equation.
    By directly predicting the wavefunction and learning from unsupervised variational Monte Carlo~\cite{Feynman_1954_theory,McMillan_1965_ground,Ceperley_1977_first,Ceperley_1980_electron,Tanatar_1989_2DEG,Foulkes_2001_review,Needs_2010_review}, neural quantum states (NQS)~\cite{Carleo_2017_solving,Nomura_2017_RBM,Carleo_2019_NetKet,Choo_2019_CNN,Sharir_2020_RNN,Pfau_2020_FermiNet,Hermann_2020_PauliNet,Spencer_2020_better,Viteritti_2023_spinformer,VonGlehn_2023_PsiFormer,Hermann_2023_review,Lange_2024_review} provide an upper bound on the ground-state energy at $\order{N^3}$ practical scaling~\cite{Pfau_2020_FermiNet}.
    A trained NQS also yields a rich set of probes, including expectation values of arbitrary observables~\cite{Foulkes_2001_review,Needs_2010_review}, for interrogating the encoded phase of matter.
    While NQS have favorable scaling compared to other methods of similar accuracy, they remain expensive to train in an absolute sense.

    Equivariant models, which are designed to exactly respect known symmetries of the problem, are widely used in other areas of physical machine learning, especially machine-learned interatomic potentials~\cite{Thomas_2018_tensor,Batzner_2022_nequip,Geiger_2022_e3nn,Liao_2023_equiformer}.
    In those tasks, equivariance improves training efficiency by sparing the model from relearning different symmetry-related views of the same underlying object.
    Condensed matter theory primarily studies periodic systems, so we wonder whether exactly enforcing translation equivariance could similarly accelerate NQS.

    We introduce TorFormer (torus transformer), a neural quantum state for the 2D electron gas which is an exact eigenstate of the total momentum for any desired target sector $\vK$.
    Using pure edge attention, TorFormer transforms input pairwise displacements into output one and two-electron features at $\order{N^2}$ cost.
    It then multiplies one-electron features by fixed plane-wave envelopes to form generalized orbitals and builds a neural Jastrow from the two-electron features.
    Envelope wavevectors are chosen automatically using the target sector $\vK$ and particle number $N$, and the model finds both the Fermi liquid and Wigner crystal without supervision.

    We test TorFormer against our own Psiformer references at $N = 36$ and $37$ and against the previous literature's best results at $N = 91$.
    At $N=36$ and $37$, TorFormer after $1\mathrm{K}$ training steps achieves lower energy and several-times-lower energy variance than Psiformer after $10\mathrm{K}$ steps.
    At $N=91$, the previous state of the art is an $800\mathrm{K}$-parameter Psiformer-based model trained for $100\mathrm{K}$ steps.
    Our $212\mathrm{K}$-parameter TorFormer already has lower energy after $1\mathrm{K}$ steps, and it continues improving over $8\mathrm{K}$ steps.
    Our improvement to the total energy at $r_s = 40.0$, excluding the trivial Madelung part, is $0.12\%$---enormous compared to the tiny differences separating phases.
    Relative to Slater-Jastrow-backflow diffusion Monte Carlo, TorFormer's energy decrease is about $9.8$ times that of the previous best NQS.
    
    We hope that our simultaneous advancements in speed and accuracy transform NQS from an expensive and specialized method into an accessible tool for accurately solving strongly correlated systems at scale.

\section{Architecture}

    \begin{figure*}[t!]
        \centering
        \includegraphics[width=0.99\linewidth]{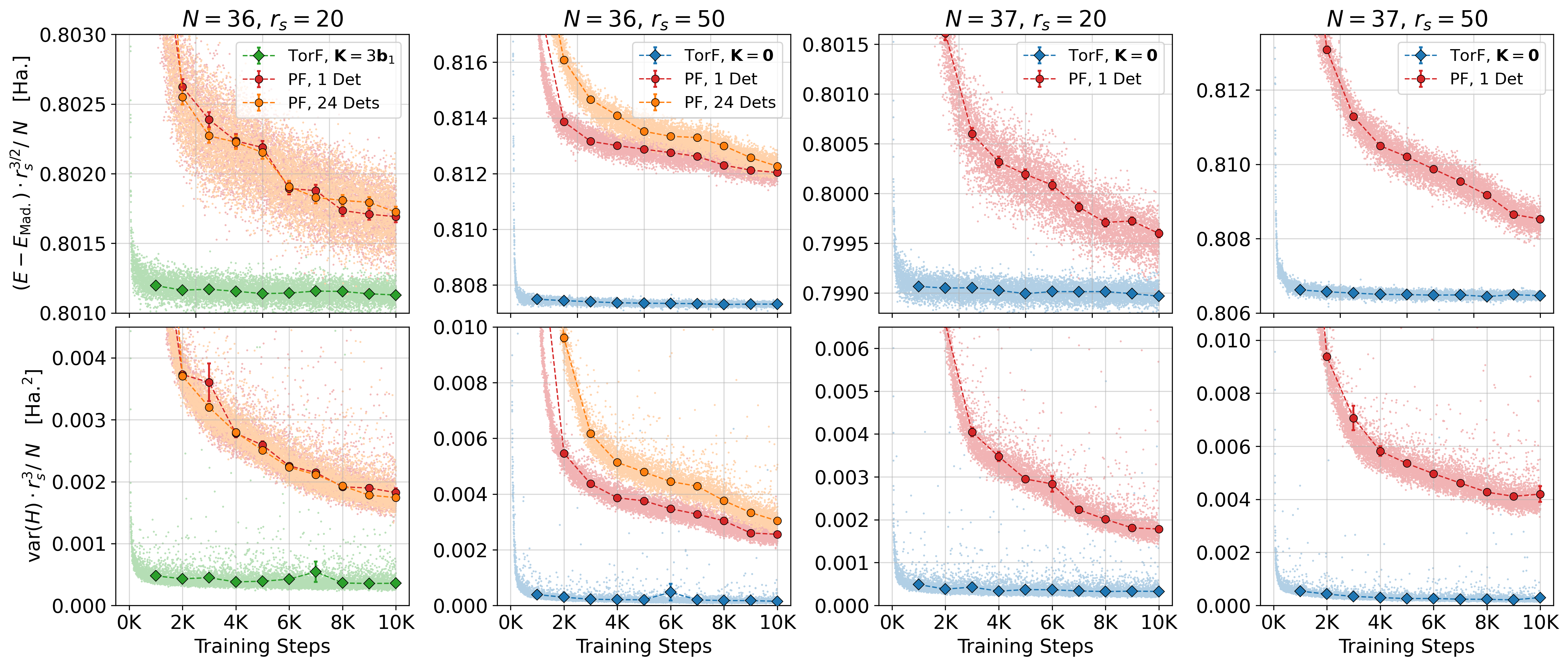}
        \caption{Medium-system comparison of TorFormer (TorF) and Psiformer (PF)'s expected energy and energy variance versus training steps. 
        Error bars every $1\mathrm{K}$ steps show inference energies estimated from fresh Markov chains, while the scatter plots show training estimates logged at each step.
        TorFormer is both faster and more accurate.}
        \label{fig:training_curves}
    \end{figure*}

    We consider $N$ spin-polarized fermions in a 2D periodic cell with direct lattice vectors $\mathbf{a}_i$ and reciprocal lattice vectors $\mathbf{b}_i$.
    The Hamiltonian is
    \begin{equation}
        H = -\frac{1}{2r_s} \sum_i \nabla^2_i + \sum_{i < j} \frac{1}{|\vr_i - \vr_j|}.
    \end{equation}
    It implicitly includes a neutralizing background and interactions with periodic images, which we treat using Ewald summation.
    We choose $\mathbf{a}_i$ so that the area per particle is $\pi$.
    We directly set the density parameter $r_s$ by scaling the kinetic energy's coefficient.
    Because the Hamiltonian commutes with translation, the total momentum $\mathbf{P} = -i \sum_i \grad_i$ is a symmetry operator.

    From a suitable neural network backbone, assume that we have output one-electron features $\vh_i$ and two-electron features $\vh_{ij}$~\footnote{Throughout, ``one-electron'' shall be synonymous with ``vertex'', and ``two-electron'' shall be synonymous with ``edge''.}.
    Because electrons are identical, features should satisfy the permutation-equivariance conditions
    \begin{equation}
        \begin{aligned}
            \vh_1 &\equiv \vh\left(\vr_1; \vr_2 \ldots \vr_N\right) = \vh\left(\vr_1; \vr_{\sigma_2} \ldots \vr_{\sigma_N}\right),\\
            \vh_{12} &\equiv \vh\left(\vr_1, \vr_2; \vr_3 \ldots \vr_N\right) = \vh\left(\vr_1, \vr_2; \vr_{\rho_3} \ldots \vr_{\rho_N}\right),
        \end{aligned}
    \end{equation}
    where $\sigma$ is a permutation of indices $2 \ldots N$, $\rho$ is a permutation of indices $3 \ldots N$, and the relations for general $\vh_i$ and $\vh_{ij}$ follow analogously.
    Additionally, the features should be invariant under a simultaneous shift of all particle coordinates. $\vh_{ij}$ is not necessarily symmetric under $i \leftrightarrow j$.
    For brevity, we will use unordered set notation $\{\vr\}$ = all particles and $\{\vr_{/i}\}$ = all particles excluding $i$.

    Within target momentum sector $\vK$ and population $N$, define the minimum-energy plane-wave fillings
    \begin{equation}
        \left\{\vk_j^\ell \mid j=1\ldots N, \ \ell=1\ldots D_{\vK, N}\right\}.
    \end{equation}
    $j$ indexes over wavevectors within each filling, $\ell$ indexes over different fillings, and $D_{\vK, N}$ is the total number of such fillings.
    $\sum_{j=1}^N \vk_j^\ell = \vec K$ for all $\ell$, and $\sum_{j=1}^N \norm{\vk_j^\ell}^2$ is the minimum possible value given the total momentum constraint.
    On average, $D_{\vK, N}$ grows slowly with $N$.
    Occasional large values reflect the genuine multireference character of open-shell Fermi systems~\cite{Mikhailov_2009_degen,Maezono_2003_sodium}.
    For the systems that we studied, $D_{\vK, N}$ was at most $24$, compared to the original FermiNet and Psiformer studies, which use either $16$ or $32$ determinants~\cite{Pfau_2020_FermiNet,VonGlehn_2023_PsiFormer}.
    
    TorFormer's readout structure is
    \begin{equation}\label{eq:torformer_readout}
        \Psi(\{\vr\}) = e^{J(\{\vr\})} \sum_{\ell} C_\ell\det_{ij} \left[e^{i \vk^\ell_j \cdot \vr_i} \varphi_{\vk^\ell_j} \left(\vr_i; \{\vr_{/i}\}\right)\right].    
    \end{equation}
    $\varphi_\vk$ are translation-invariant generalized orbitals built from $\vh_i$, $J$ is a translation-invariant neural Jastrow built from $\vh_{ij}$, and $C_\ell$ is a trainable but position-independent coefficient.
    If we shift all particle coordinates $\vr_i \rightarrow \vr_i+\Delta\vr$, nothing in Eq.~\ref{eq:torformer_readout} changes except for the fixed plane-wave envelopes, which yield $\Psi(\{\vr + \Delta\vr\}) = e^{i\vK \cdot \Delta \vr}\Psi(\{\vr\})$.
    
    Our envelope-based architecture is more flexible than backflow~\footnote{By backflow, we mean the canonical construction where raw single-particle coordinates $\vr_i$ are shifted by permutation-equivariant displacements before being plugged in to mean-field single-particle orbitals.}, because plane waves can receive separate learned modifications instead of sharing a common backflow displacement.
    Previous work proved that backflow around plane waves is universal for many-body ground states~\cite{Pescia_2024_MPNQS}, so TorFormer is universal as well.

    All determinants share a common pool of orbitals $\varphi_\vk$.
    We force $\varphi_\vk = -i\varphi_{-\vk}^*$ for $\vk \neq \mathbf{0}$, and we force the special zero-wavevector orbital $\varphi_\mathbf{0}$ to be real.
    We parameterize the free orbitals as
    \begin{equation}
        \begin{aligned}
            \varphi_{\vk \neq \mathbf{0}}\left(\vr_i; \{\vr_{/i}\}\right) &= \left(\mathbf{w}_\vk^\text{Re} + i \mathbf{w}_\vk^\text{Im}\right) \cdot \vh_i + \left(\mathbf{b}_\mathbf{k}^\mathrm{Re} + i\mathbf{b}_\mathbf{k}^\mathrm{Im}\right),\\
            \varphi_{\vk = \mathbf{0}}\left(\vr_i; \{\vr_{/i}\}\right) &= \mathbf{w}_\mathbf{0}^\mathrm{Re} \cdot \vh_i + \mathbf{b}_\mathbf{0}^\mathrm{Re}.
        \end{aligned}
    \end{equation}
    If a filling $\{\vk_j^\ell \mid j=1\ldots N \}$ maps back to itself under inversion $\vk \rightarrow -\vk$, we force its coefficient $C_\ell$ to be real; otherwise $C_\ell$ is complex.
    All of these constraints are motivated by the fact that a zero-momentum ground state can be made purely real.
    Specifically, our parameterization guarantees that for closed-shell $N$, the wavefunction is automatically real.
    For non-closed-shell $N$ at $\vK = \mathbf{0}$, we take the real part at the end before returning $\Psi$.

    We parameterize the Jastrow as $J = J_\mathrm{N}  + J_\mathrm{M}$. $J_\mathrm{N}$ is
    \begin{equation}
        J_\mathrm{N}(\{\vr\}) = \sum_{i \neq j} \left( \mathbf{w}_\mathrm{J} \cdot \vh_{ij} \ e^{1 - \sqrt{1 + \left[\ln(1 + e^\gamma) \left|\vr_i - \vr_j\right|_\mathbf{a}\right]^2}} \right),
    \end{equation}
    where $|\cdot|_\mathbf{a}$ is the periodified distance from Ref.~\cite{Cassella_2023_wigner}.
    Because the sum is over $\order{N^2}$ terms compared to the $\order{N}$ scaling expected of the log wavefunction, we apply a softened exponential damping.
    The short-range softening keeps the Jastrow smooth, while the softplus on the decay parameter $\gamma$ ensures positive decay rate.
    $J_\mathrm{M}$ is a Psiformer-style ``manual'' Jastrow factor, which exactly enforces the electron-electron Kato cusp.
    The rest of our model is smooth in the electron coordinates.

    To generate $\vh_i$ and $\vh_{ij}$ satisfying the aforementioned conditions, we use the EVE (edge-vertex-edge) backbone introduced by our previous work~\cite{Dai_2026_EVE} on bosonic neural momentum eigenstates, up to a few small changes.
    We zero-initialize the edge attention's logit kernels (Ref.~\cite{Dai_2026_EVE}, Eq. 6), so initial behavior is mean pool.
    We also zero-initialize the part of $\MLP_1$'s second linear layer that generates the vertex-to-edge perturbations $\vec y^\mathrm{row}$ and $\vec y^\mathrm{col}$ (Ref.~\cite{Dai_2026_EVE}, Eq. 7), so initially information only flows from edges to vertices.
    Finally, we made the manual Jastrow's $\alpha$ parameter (Ref.~\cite{VonGlehn_2023_PsiFormer}, Eq. 9) and the featurizer's $\alpha_\mathrm{soft}$ (Ref.~\cite{Dai_2026_EVE}, Eq. 8) trainable.
    
    We have excellent priors for the ground-state momentum sector in most cases.
    Here, we derive selection rules for a few common situations.

    \begin{figure}[t!]
        \centering
        \includegraphics[width=0.98\linewidth]{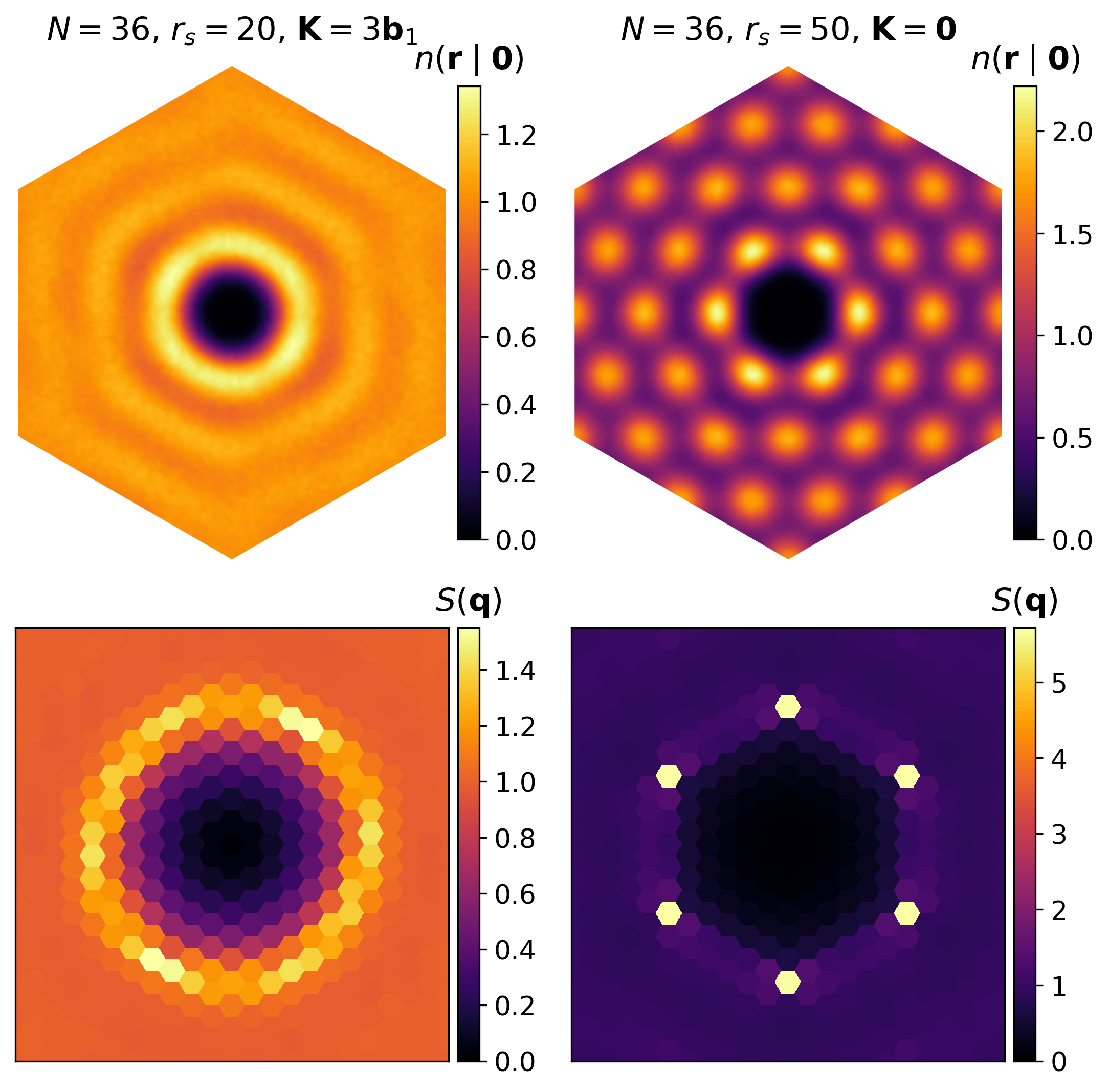}
        \caption{Pair correlation function (top row) and static structure factor (bottom row) for $r_s = 20.0$ (left column) and $r_s = 50.0$ (right column). The expected phase is apparent in both observables.}
        \label{fig:correlations_n36}
    \end{figure}
    
    Commensurate triangular-lattice Wigner crystals are specified by indices $(m, n)$ and contain $N = m^2 + mn + n^2$ particles.
    Defining the crystal's direct lattice vectors $\Tilde{\vec{a}}_1$ and $\Tilde{\vec{a}}_2$, the simulation cell's lattice vectors are $\vec{a}_1 = m \Tilde{\vec{a}}_1 + n \Tilde{\vec{a}}_2$ and $\vec{a}_2 = -n \Tilde{\vec{a}}_1 + (m + n) \Tilde{\vec{a}}_2$.
    Translations by $\Tilde{\vec{a}}_1$ and $\Tilde{\vec{a}}_2$ are equivalent to permutations of the lattice sites.
    Some basic modular arithmetic shows these permutations can be decomposed into $N - \gcd(m, n)$ swaps.
    Iterating through the $4$ possibilities for $m$ and $n$'s parities shows that $N - \gcd(m, n)$ is always even.
    Thus, a pinned Wigner crystal has nonzero amplitude only in total momentum sectors lying in the crystal's reciprocal lattice.
    We expect the zero-momentum floating crystal to minimize center-of-mass kinetic energy and be the ground state.

    On the fluid side, we choose candidate momentum sectors by examining the free-particle problem.
    For closed-shell $N$, the free-particle ground state is unique and has zero total momentum from inversion symmetry $\vk \leftrightarrow -\vk$.
    Below the crystallization transition, the ground state should be adiabatically connected to the noninteracting limit, so we expect $\vK = \mathbf{0}$.
    For open-shell $N$, we can do perturbation theory to generate a few promising sectors and just test them.
    In practice, most works use magic numbers which are both closed-shell and fit a commensurate crystal, which trivializes sector selection.

    \begin{figure}[t!]
        \centering
        \includegraphics[width=0.98\linewidth]{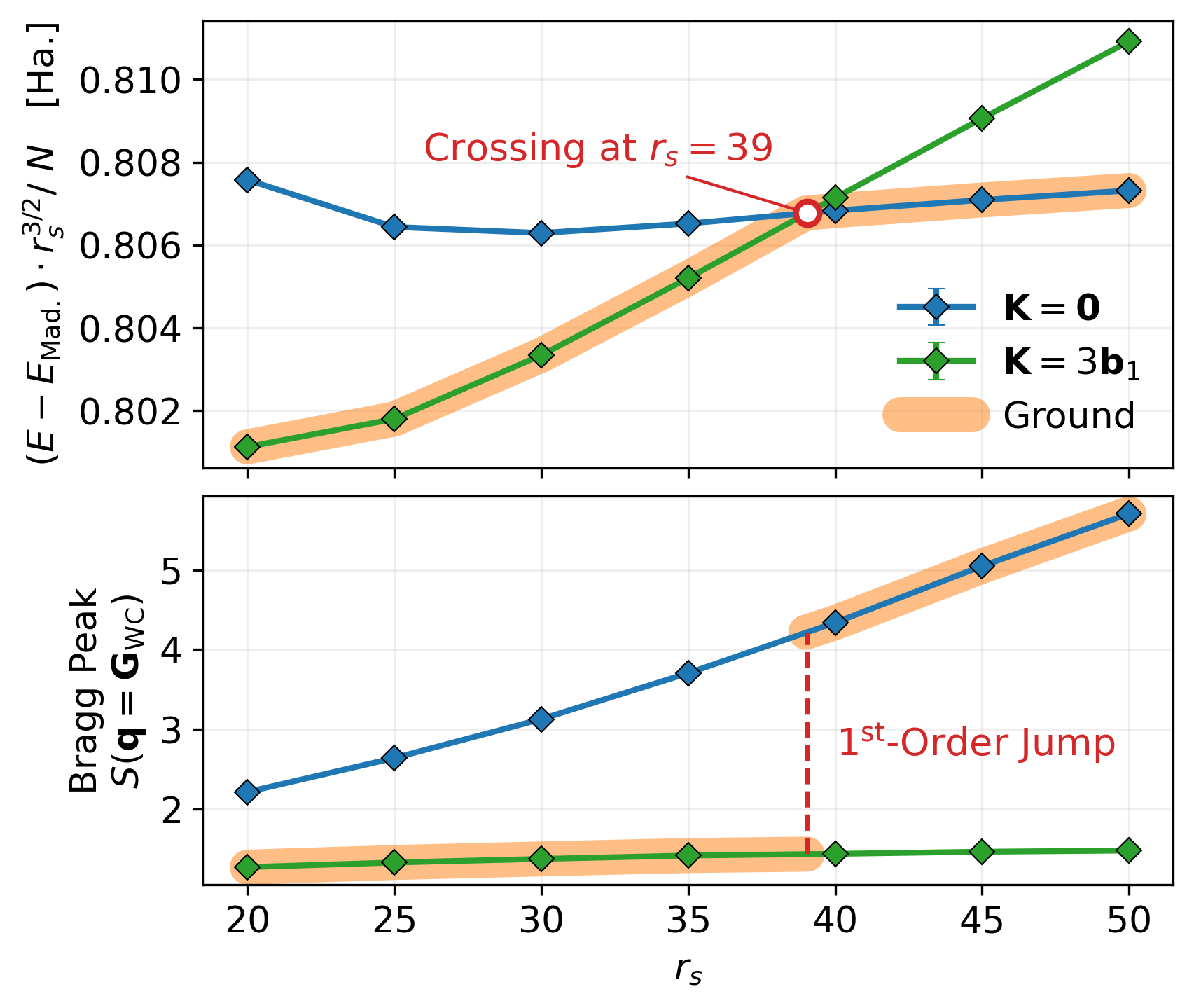}
        \caption{TorFormer's energy per particle and Bragg peak height for sectors $\vK = 3 \mathbf{b}_1$ and $\vK = \mathbf{0}$, plotted as a function of density parameter $r_s$.
        The symmetry-protected crossing at $r_s = 39$ causes a jump in the static structure factor, which is a finite-size realization of the phase transition.}
        \label{fig:sector_crossing}
    \end{figure}

\section{Results}

    All systems use triangular-lattice simulation cells, with lattice vectors $\vec{a}_1$ and $\vec{a}_2$ of equal length and separated by an angle of $60^\circ$.
    Our TorFormer models have one-electron dimension $d_1 = 128$, two-electron dimension $d_2 = 32$, $4$ edge-attention heads, and $4$ layers, giving about $200\mathrm{K}$ parameters.
    Our Psiformer reference has the same model dimension, number of heads, and number of layers as the original ``small Psiformer''~\cite{VonGlehn_2023_PsiFormer} and has $1.6\mathrm{M}$ to $2\mathrm{M}$ parameters.
    We match the number of determinants in Psiformer to the number of fillings that TorFormer uses.

    Although translation-equivariance is strictly speaking not the only difference between TorFormer and Psiformer, it informs almost every design decision in TorFormer. For example, the EVE backbone was specifically designed to output translation-invariant features, which is unnatural for other backbones such as self-attention.

    Our experiments used eight V100 GPUs, four L40S GPUs, or four H200 GPUs.
    Except for the H200s, which we only used for one $N=91$ run, these resources are relatively accessible at many research organizations.
    On our hardware at $N = 36$ and $37$, the two models have comparable time per training step.
    Training uses a batch size of $2048$ and $10\mathrm{K}$ steps for the medium-scale experiments and $8\mathrm{K}$ steps for the large-scale ones.

    \begin{figure}[t!]
        \centering
        \includegraphics[width=0.98\linewidth]{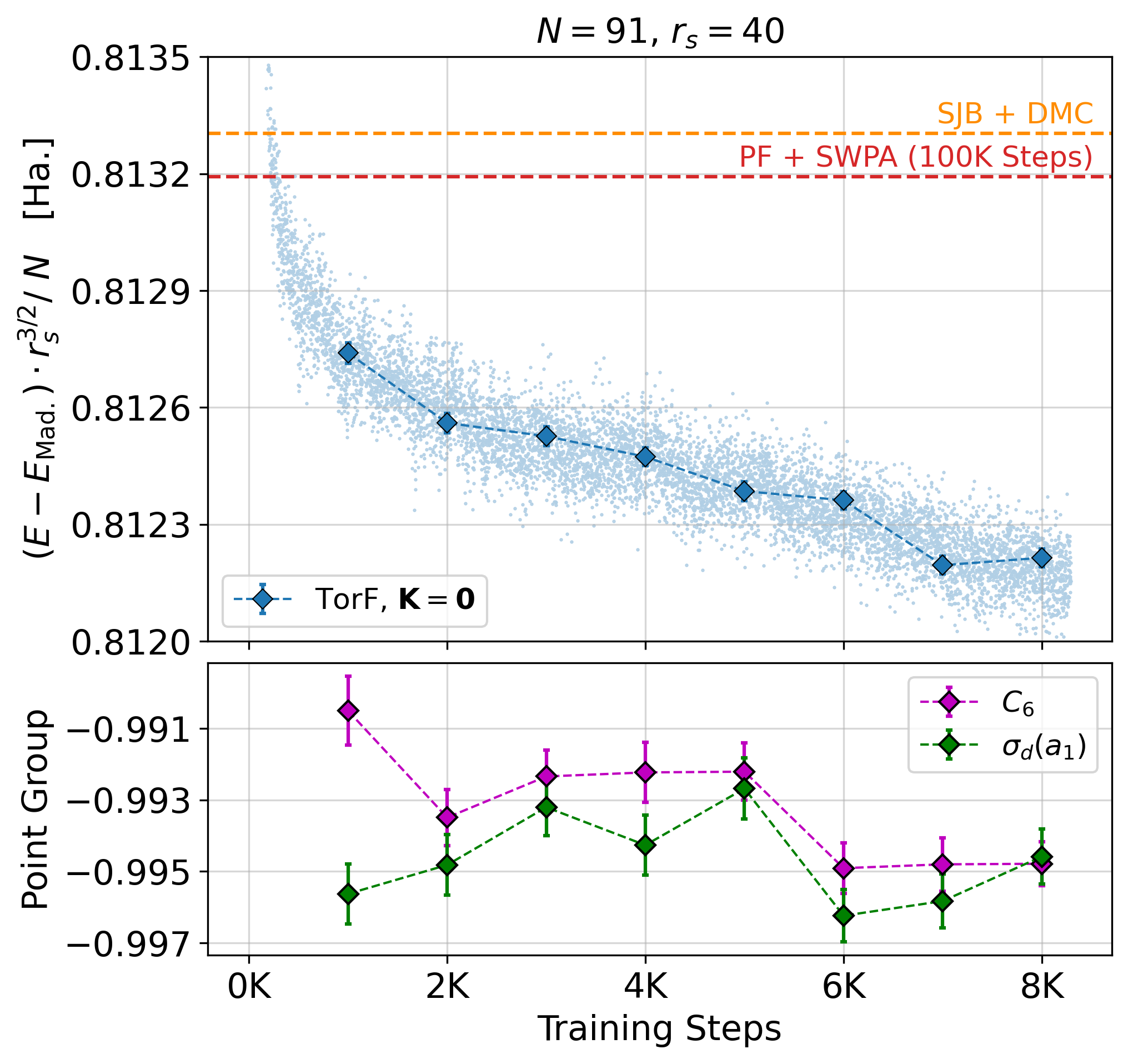}
        \caption{$N = 91$, $r_s = 40.0$ training curves. Top panel: energy comparison between TorFormer, the previous best NQS (Psiformer+SWPA), and Slater-Jastrow-backflow diffusion Monte Carlo. TorFormer is more accurate than Psiformer+SWPA while using less than a tenth of the training steps. Bottom: TorFormer is a near-exact eigenstate of the triangular lattice point-group generators with eigenvalue $-1$ for both.}
        \label{fig:n91_training_curves}
    \end{figure}   
       
    \subsection{Medium-Scale Experiments at $N=36$ and $37$}

        $N = 36$ is commensurate with a $6 \times 6$ Wigner crystal, and it is one particle below the closed shell $N = 37$.
        This forces the Fermi liquid to be six-fold degenerate over sectors $\pm 3\mathbf{b}_1$, $\pm3\mathbf{b}_2$, and $\pm3(\mathbf{b}_1 + \mathbf{b}_2)$.
        We test TorFormer at $\vK = 3\mathbf{b}_1$, where there is a single minimum-energy plane-wave filling, and $\vK = \mathbf{0}$, where there are $24$ fillings.
        To match TorFormer's determinant count, we run $1$- and $24$-determinant Psiformer references.
    
        $N = 37$ is a magic number, being both closed-shell and fitting a $(4, 3)$ Wigner crystal and its reflection partner $(3, 4)$.
        Because the ground state has zero momentum and is unique, it is also an eigenstate of the hexagonal point group $D_6$.
        We find that the Fermi liquid and Wigner crystal lie in the same space group sector, so the transition manifests as a smooth avoided crossing.
        Our Psiformer reference here uses $1$ determinant.
        
        Fig.~\ref{fig:training_curves} shows the expected energy and energy variance as a function of training steps for TorFormer and Psiformer on $N=36$ and $37$ at $r_s = 20.0$ and $50.0$.
        For the expected energy, we subtract the trivial Madelung part, rescale by the semiclassical zero-point energy scale, and report energy per-particle.
        For the energy variance, we apply the same rescaling and report variance-per-particle.
        At only $1\mathrm{K}$ steps, TorFormer is several times more accurate (based on energy variance, which is empirically proportional to energy error~\cite{Fu_2024_variance}) than Psiformer at $1\mathrm{K}$.
        Psiformer typically requires on the order of $100\mathrm{K}$ steps for full convergence~\cite{VonGlehn_2023_PsiFormer}.
        TorFormer's ability to converge in only a few thousand steps therefore represents a more than order-of-magnitude improvement.
    
        \begin{figure}[t!]
            \centering
            \includegraphics[width=0.98\linewidth]{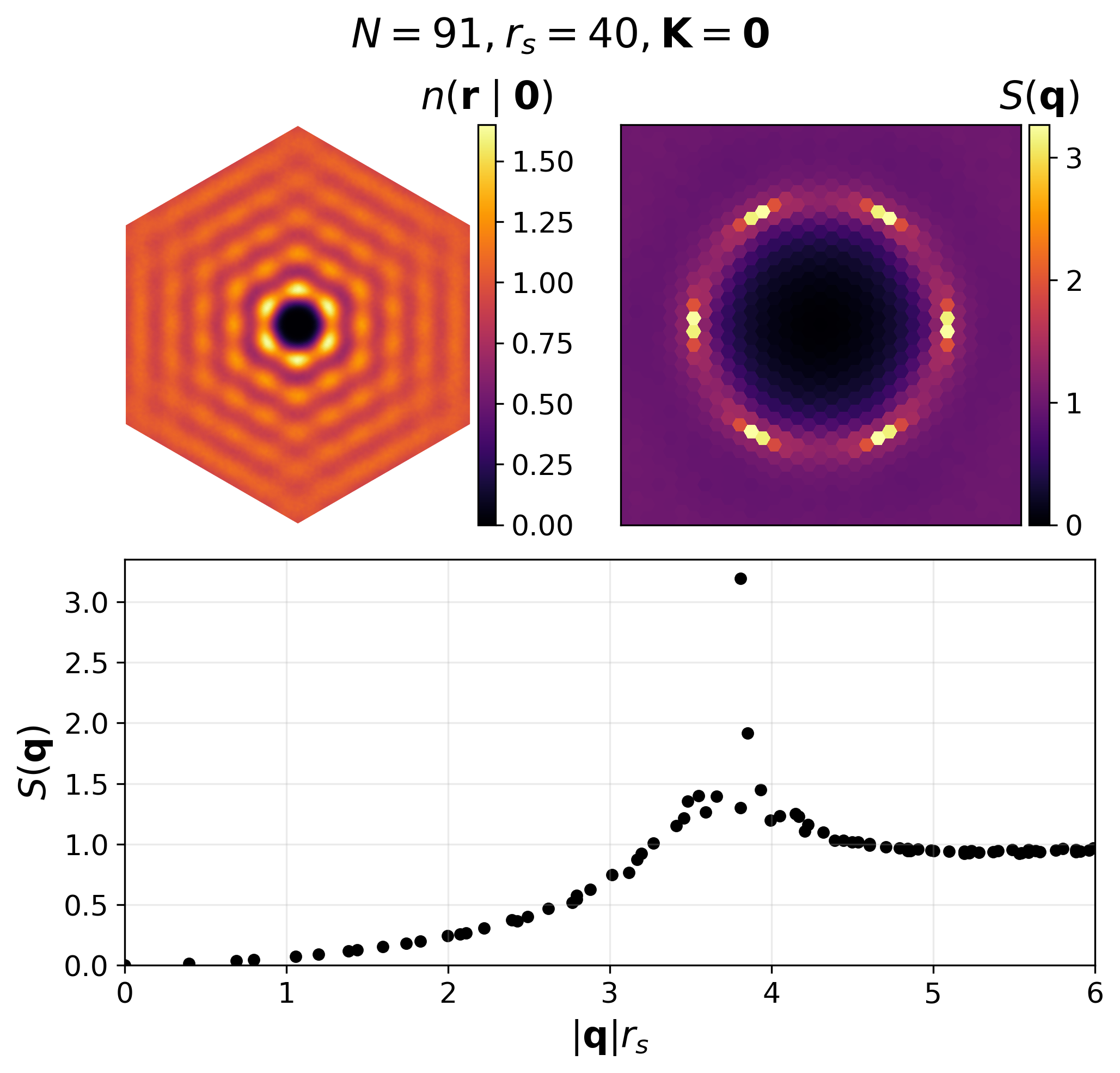}
            \caption{$N = 91$, $r_s = 40.0$ correlations.
            Top left: pair correlation.
            The pattern is smeared, because the model learned a superposition of the $(6, 5)$ and $(5, 6)$ crystal embeddings.
            Top right: 2D map of static structure factor.
            Each of the two crystal embeddings contributes a set of six Bragg peaks for twelve total.
            Bottom: symmetry-averaged static structure factor versus wavevector norm, showing the Bragg peak.}
            \label{fig:n91_correlations}
        \end{figure}
    
        Fig.~\ref{fig:correlations_n36} shows the pair correlation function (equivalent to a histogram of $\vr_i - \vr_j$) and the static structure factor $S(\vec{q}) = \frac{1}{N}\left\langle \sum_{i,j} e^{i \vec{q} \cdot (\vr_i - \vr_j)} \right\rangle$ for $N = 36$.
        For $r_s = 20.0$, we see the features expected of a liquid.
        The pair correlation has a hole at short range with faint ringing farther away, while $S(\vec q)$ lacks any Bragg peaks.
        Slight anisotropy is expected because the sector $\vK = 3\mathbf{b}_1$ picks out a direction.
        At $r_s = 50.0$, we see periodic peaks in the pair correlation forming a triangular lattice, and $S(\vec q)$ shows the expected six-fold peaks at the crystal's first reciprocal-lattice shell.
        Thus, TorFormer learned both the Fermi liquid and Wigner crystal without supervision.
    
        Fig.~\ref{fig:sector_crossing} shows that $N = 36$ exhibits a symmetry protected crossing, which is a finite-size realization of the crystallization transition.
        Here, $3\mathbf{b}_1$ and $\mathbf{0}$ correspond to the Fermi liquid and Wigner crystal, although the target sectors themselves do not force the model to represent one phase or the other.
        From $S(\vec q)$, the liquid survives as an excited state for $r_s$ above the transition, and vice versa for the Wigner crystal. It is much harder for non-equivariant models such as Psiformer to target these excited states.
        For $N = 37$, we found that the two phases lie in the same space group sector and smoothly cross into each other over a wide $r_s$ window (not shown in figure).
        VMC using separate hand-crafted ans\"atze for the liquid and crystal generically predicts a first-order crossing in all cases.
        In contrast, our unified approach correctly learns either the sharp or smooth behavior depending on $N$.

    \subsection{Large-Scale Experiments at $N = 91$}

\newlength{\energyresultwidth}
\settowidth{\energyresultwidth}{\num{0.810485(5)}}

\newcommand{\energyresult}[1]{%
    \makebox[\energyresultwidth][l]{\num{#1}}%
}
\newcommand{\boldenergyresult}[1]{%
    \makebox[\energyresultwidth][l]{\bfseries\num{#1}}%
}

\newlength{\reductionresultwidth}
\settowidth{\reductionresultwidth}{\num{0.014}\%}

\newcommand{\reductionresult}[1]{%
    \makebox[\reductionresultwidth][l]{\num{#1}\%}%
}
\newcommand{\boldreductionresult}[1]{%
    \makebox[\reductionresultwidth][l]{\bfseries\num{#1}\%}%
}
\newcommand{\noreduction}{%
    \makebox[\reductionresultwidth][c]{$\cdots$}%
}

\begin{table}[t]
    \centering
    \begin{tabular}{
        c@{\hspace{5pt}}|@{\hspace{8pt}}
        c@{\hspace{5pt}}|@{\hspace{8pt}}
        c@{\hspace{5pt}}|@{\hspace{8pt}}
        c
    }
        \hline
        System
        & Method
        & $\frac{(E-E_\mathrm{Mad.})r_s^{3/2}}{N}$ [Ha.]
        & Reduction \\
        \hline

        $N = 91$
        & SJB+DMC
        & \energyresult{0.81059(2)}
        & \noreduction \\

        $r_s = 30$
        & PF+SWPA
        & \energyresult{0.810485(5)}
        & \reductionresult{0.013} \\

        &
        \textbf{TorFormer}
        & \boldenergyresult{0.80962(2)}
        & \boldreductionresult{0.11} \\
        \hline

        $N = 91$
        & SJB+DMC
        & \energyresult{0.81330(5)}
        & \noreduction \\

        $r_s = 40$
        & PF+SWPA
        & \energyresult{0.813192(5)}
        & \reductionresult{0.014} \\

        &
        \textbf{TorFormer}
        & \boldenergyresult{0.81221(2)}
        & \boldreductionresult{0.12} \\
        \hline
    \end{tabular}

    \caption{$N = 91$ comparison of TorFormer (trained for only $8\mathrm{K}$ steps); Psiformer+SWPA, the previous best NQS (trained for $100\mathrm{K}$ steps)~\cite{Gaggioli_2026_SWPA}; and Slater-Jastrow-backflow diffusion Monte Carlo (DMC)~\cite{Gaggioli_2026_SWPA}. TorFormer achieves a significantly better energy than Psiformer+SWPA using $<10\%$ the steps. We measure reduction against the previous best method.}
    \label{tab:n91_comparison}
\end{table}
        
        We test TorFormer at $N = 91$ because high-quality NQS results are available in the literature~\cite{Gaggioli_2026_SWPA}.
        Comparing against external benchmarks rather than our own also removes a potential source of human bias.
        $N = 91$ is both closed-shell and fits commensurate Wigner crystals with indices $(6, 5)$ and $(5, 6)$, so we run TorFormer at $\vK = \mathbf{0}$.
        As for $N = 37$, both the Fermi liquid and Wigner crystal are eigenstates of the point group operators and lie in the same sector.
        The previous work also found that $N = 91$ was sufficient to nearly eliminate finite-size error~\cite{Gaggioli_2026_SWPA}.
        
        To the best of our knowledge, the current state of the art at $N = 91$ is a Psiformer-like model~\cite{Gaggioli_2026_SWPA}.
        It improves upon the original Psiformer by adding a distance bias $-r_{ij}/\lambda$ to the attention logits~\cite{Gaggioli_2026_SWPA}.
        We refer to this model as Psiformer+SWPA (spatially weighted particle attention) following the authors' terminology.
        At time of writing, the authors only specify that their model has $800\mathrm{K}$ parameters and was trained for $100\mathrm{K}$ steps, so we compare based on number of training steps~\cite{Gaggioli_2026_SWPA}.
        Differing per-step time is very unlikely to meaningfully change the conclusions given the size of our gains.

        Fig.~\ref{fig:n91_training_curves}'s top panel shows that TorFormer surpasses Psiformer+SWPA after only a few hundred iterations and continues improving through the entire training horizon.
        Crucially, the time to surpass Psiformer was also a few hundred iterations at $N = 36$ and $37$, indicating that TorFormer's gains are approximately consistent across system size~\footnote{TorFormer takes longer to surpass the $N = 91$ Psiformer reference, but this is expected given the reference's longer training horizon at $100\mathrm{K}$ compared to $10\mathrm{K}$ (and the fact that the reference is improved via SWPA).}.
        Table~\ref{tab:n91_comparison} compares final inference energies for TorFormer, Psiformer+SWPA, and Slater-Jastrow-backflow diffusion Monte Carlo (also from Ref.~\cite{Gaggioli_2026_SWPA}).
        The data show that TorFormer is a significant improvement over both Psiformer+SWPA and SJB-DMC.
        In contrast, Psiformer+SWPA only improves modestly over SJB-DMC.
        At $r_s = 40.0$, TorFormer's final energy is $0.12\%$ lower than Psiformer+SWPA, which is enormous compared to the tiny differences separating phases.
        
        Fig.~\ref{fig:n91_training_curves}'s bottom panel shows that TorFormer respects the point group symmetries almost exactly, despite us only explicitly enforcing translation symmetry.
        Fig.~\ref{fig:n91_correlations}'s $S(\vec q)$ plot shows that TorFormer learned a cat state superposing the $(6, 5)$ and $(5, 6)$ Wigner crystals.
        Because the two crystal embeddings' wavefunctions are peaked in different parts of configuration space, the expectation value of an observable that is diagonal in the position basis is approximately the average of the expectations in each orientation individually.        
        $S(\vec q)$ follows this picture, with each orientation contributing six peaks for a total of twelve, all at the expected locations.
        The pair correlation is inconclusive but not inconsistent with the cat state, because superimposing the $(6, 5)$ and $(5, 6)$ patterns causes significant smearing.

\section{Discussion}

    Our work demonstrates that building translation-equivariance into a neural quantum state dramatically improves both training speed and final energy on the 2D electron gas up to $N = 91$ near the thermodynamic limit~\cite{Gaggioli_2026_SWPA}.
    While our current work focuses on improving results for the global ground state, in the future it would be interesting to apply TorFormer to solve for arbitrary-$\mathbf{K}$ Fermi ground states (i.e. quasiparticles), similar to what our previous EVE model did for bosons~\cite{Dai_2026_EVE}.

    Sufficient \textit{quantitative} efficiency gains will change the \textit{qualitative} role that NQS play in research.
    A calculation requiring weeks on a large GPU cluster must be pre-planned as a major endeavor, and its long turnaround time discourages rapid experimentation with either the model or the Hamiltonian.
    In contrast, a calculation that completes in hours to a few days on a handful of GPUs becomes part of the everyday research cycle---launched spontaneously, iterated on rapidly, and supported by standard institutional allocations.
    Our long-term vision is for NQS to take the role that Hartree–Fock plays today—a routine and widely accessible front-line method, only vastly more capable and flexible.
    TorFormer is the first step toward that goal.
    
\section*{Acknowledgments}
    We thank Yuta Sakai, Kai Tak Lam and Jeff Wu for insightful discussions.
    DDD was supported by the National Science Foundation Graduate Research Fellowship under Grant No. 2141064.
    The authors acknowledge the MIT SuperCloud, Lincoln Laboratory Supercomputing Center (LLSC), and MIT Office of Research Computing and Data (ORCD) for providing high-performance computing resources.
    This material is based upon work supported in part by the MIT Institute for Soldier Nanotechnologies. 
    This work is supported by the National Science Foundation under Cooperative Agreement PHY-2019786 (The NSF AI Institute for Artificial Intelligence and Fundamental Interactions, http://iaifi.org/).
    AI agents were used to support programming tasks.
    The authors wrote all prose in this paper by hand.

\bibliography{references}

\end{document}